\documentclass[12pt,a4]{article}
\usepackage{relsize}
\usepackage{array}
\usepackage{a4wide}
\usepackage{enumerate}
  \usepackage[linktocpage=true]{hyperref} 
\usepackage{amssymb,euscript,amsmath,amsfonts,filecontents}
\usepackage{mathtools} 
\usepackage[utf8]{inputenc}
\usepackage{enumerate}
\usepackage{hypernat}
\usepackage{mathtools} 
 
\begin{document}

\renewcommand{\thefootnote}{\fnsymbol{footnote}}
 
 \centerline{\LARGE \bf {\sc Non-Lorentzian Avatars} }
 \medskip
\centerline{\LARGE \bf {\sc  of   (1,0) Theories}}
 
 \vspace{1cm}
 
   \centerline{
   {\large {\bf   {\sc N.~Lambert${}^{\,a,b}$}}\footnote{E-mail address: \href{neil.lambert@kcl.ac.uk}{\tt neil.lambert@kcl.ac.uk}  }  \,   {\sc    and T.~Orchard${}^{\,a}$}}\footnote{E-mail address: \href{mailto:tristan.orchard@kcl.ac.uk}{\tt tristan.orchard@kcl.ac.uk}} }

\vskip 1cm
\centerline{${}^a${\it Department of Mathematics}}
\centerline{{\it King's College London, WC2R 2LS, UK}}

\vspace{1.0truecm}

 
\thispagestyle{empty}

\centerline{\sc Abstract}
We construct five-dimensional non-Lorentzian Lagrangian gauge field theories with an $SU(1,3)$ conformal symmetry and 12 (conformal) supersymmetries. Such theories  are interesting in their own right but  can arise from  six-dimensional (1,0) superconformal field theories on a conformally compactified Minkowski spacetime. In the limit that the conformal compactification is removed the Lagrangians we find give field theory formulations of  DLCQ constructions  of six-dimensional $(1,0)$ conformal field theories.     

\newpage

\section{Introduction}

Starting with \cite{Seiberg:1996qx}, there is now known to be a wide variety of six-dimensional $(1,0)$ superconformal field theories, for example see \cite{Heckman:2015bfa,Bhardwaj:2015xxa}. As with their $(2,0)$ cousins \cite{Witten:1995zh}, these do not readily possess Lagrangian descriptions, although they typically reduce to five-dimensional super-Yang-Mills theories upon compactification on $S^1$. In fact, a class of six-dimensional Lagrangians with $(1,0)$ supersymmetry has been constructed in \cite{Samtleben:2011fj} (see also \cite{Chen:2017kgl} for an on-shell construction)\footnote{And   \cite{Lambert:2019diy} for a $(2,0)$ Lagrangian theory with $SU(2)$ gauge group.}.  However, we generically see that Lagrangian field theories of Lorentzian signature above four dimensions are non-renormalisable, or have  unbounded energy from below. It is thus not clear what relationship such classical actions have with their strongly coupled quantum counterparts.   
In this paper we will construct a family of five-dimensional non-Lorentzian Lagrangian theories, with 12 (conformal) supersymmetries, and an $SU(1,3)$ spacetime symmetry. We believe this provides a novel class of higher dimensional theories worthy of further exploration. 

There is a growing interest in non-Lorentzian field theories from a variety of view points and applications, for some recent studies see \cite{Karch:2020yuy,Cremonini:2020rdx,Bergshoeff:2020baa,Blair:2020ops,Kluson:2019avy,Lambert:2018lgt,Harmark:2018cdl,Festuccia:2016caf,Golkar:2014mwa}. Our own interest stems from a novel form of conformal compactification, which appears to be able to reproduce aspects of the original non-compact six-dimensional SCFT from a five-dimensional Lagrangian theory with a Kaluza-Klein-like tower \cite{Lambert:2019jwi,Lambert:2019fne,Lambert:2020zdc}. In particular, the mode number of the tower is identified with the charge arising from the topological current \begin{equation}
J = \frac1{8\pi^2}\star {\rm tr} (F\wedge F)	\ .
\end{equation}

Reduction of a six-dimensional Lorentzian field theory can lead to five-dimensional non-Lorentzian field theories, if the compact direction is taken to be null. This breaks the $SO(2,6)$ conformal group in six-dimensions to an $SU(1,3)$ symmetry. Thus the Lagrangians of this paper have the correct symmetries to be identified with the null conformal compactification of six-dimensional superconformal field theories. In the limit that the conformal compactification is removed, these Lagrangians have on-shell constraints that reduce the dynamics to motion on instanton moduli space. These constraints arise naturally from the imposition of a null isometry, and their realisation is achieved without adding additional unphysical degrees of freedom in the form of Lagrange multipliers. In fact, different components of the fields in six-dimensions act as Lagrange multipliers to  other fields in five dimensions. In this way, in the limit where there is no conformal compactification, we recover the DLCQ prescriptions for six-dimensional superconformal field theories \cite{Aharony:1997pm,Aharony:1997an}. It is therefore our hope that the Lagrangians presented here can be used to understanding six-dimensional (1,0) theories.

The rest of this paper is organised as follows. In section two we briefly discuss the conformal compactification of Minkowski space leading to so-called $\Omega$-deformed Minkowski space. In section three we discuss $(1,0)$ conformal supermultiplets and their null reduction. In section four we present a class of non-Lorentzian gauge theories in 4+1 dimensions with 4(+8) (conformal)supersymmetries, and an $SU(1,3)$ conformal symmetry. Section five contains our discussion and conclusions. We also include an appendix which gives in detail the relation of the six-dimensional spinors used here to those of eleven-dimensions.

  \section{Conformal Field Theories on $\Omega$-deformed Minkowski Space}

We start by observing that the metric of six-dimensional Minkowski space can be written as
\begin{equation}
\begin{split}\label{ConfTran}
d\hat s^2_{Minkowski} &= -2d\hat x^+d\hat x^- + \delta_{ij}d\hat x^id\hat x^j\\
&=\frac{1}{\cos^2(x^+/2R)} d\hat s^2_\Omega 	\ ,
\end{split}
  \end{equation}
where
  \begin{align}\label{OM}
d\hat s^2_\Omega = -2dx^+(dx^- - \frac12 \Omega_{ij}x^idx^j) + \delta_{ij}dx^idx^j\ .	
  \end{align}
  Here $\Omega = -\star \Omega$ and $\Omega_{ik}\Omega_{kj} = -R^{-2}\delta_{ij}$.  We refer to (\ref{OM}) as $\Omega$-deformed Minkowski space. Clearly for $R\to\infty$ we recover ordinary Minkowski spacetime.  Furthermore, (\ref{OM}) naturally arises  as the boundary of  $AdS_7$ in the parametrisation of \cite{Pope:1999xg}.

  The conformal transformation (\ref{ConfTran}) maps the $\hat x^+\in\mathbb R$ coordinate of Minkowski space into a finite range $x^+\in (-\pi R,\pi R)$. Although the fields need not be periodic in $x^+$, their restriction to a finite range is sufficient to capture the entire dynamics of the non-compact Minkowski spacetime.  Thus a superconformal field theory on six-dimensional Minkowski space is equivalent the same theory on $\Omega$-deformed Minkowski space, with $x^+\in (-\pi R,\pi R)$. By expanding as a Fourier series in $x^+$, the full non-compact six-dimensional theory can be realised by a discrete Kaluza-Klein tower of five-dimensional operators which depend on $(x^-,x^i)$.
  
   It also follows that (\ref{OM}) admits the maximal number of conformal Killing spinors and Killing vectors. Although some will have explicit $x^+$ dependence and hence won't survive a reduction to the zero-Fourier mode sector. Such generators will not lead to symmetries of the five-dimensional Lagrangian  constructed from the zero-modes, but could nevertheless still be present in the quantum theory.
  It was shown in  \cite{Lambert:2019jwi} that the reduction to the zero-modes preserves, classically, 3/4 of the supersymmetries and conformal supersymmetries. In addition in \cite{Lambert:2019fne} it was shown that the reduction preserves a $SU(1,3)$ conformal symmetry. The $SU(1,3)$ conformal symmetry  imposes non-trivial restrictions on the $n$-point functions \cite{Lambert:2020zdc}.
  
  Furthermore, the Fourier mode number is naturally identified with the instanton number of the five-dimensional gauge theory. This raises the possibility of reconstructing correlation functions of the full six-dimensional theory from  the five-dimensional Lagrangian gauge theory \cite{Lambert:2020zdc}. It is therefore natural to construct more general  $(1,0)$ supersymmetric conformal field theories, obtained by reduction on Omega-deformed Minkowski space.

 We can also consider the case $\Omega_{ij}=0$. Here the metric is six-dimensional Minkowski space where the range of  $x^+$ is the whole real line. Compactifying $x^+$ corresponds to the familiar DLCQ matrix model proposals for $(1,0)$ and $(2,0)$  theories  \cite{Aharony:1997pm,Aharony:1997an},  where the dynamics is  described by quantum mechanics on the  moduli space of instantons.

\section{Six-Dimensional Conformal Multiplets and their Reduction}

There are known to be interacting superconformal field theories in six-dimensions with $(1,0)$ supersymmetry. However, it is not thought that a Lagrangian description for such theories exists. Nevertheless, if we allow ourselves to think in terms of fields there are several types of linear six-dimensional $(1,0)$ supermultiplets: vector-multiplets,  linear-multiplets, scalar-multiplets,  hyper-multiplets,  tensor-multiplets and half-hyper-multiplets \cite{Bergshoeff:1985mz,Ferrara:2000xg} (see \cite{Buican:2016hpb,Cordova:2016emh} for a detailed classification).

In this paper we will not consider vector and linear multiplets, as their scaling dimensions are incompatible with a quadratic kinetic term, it could be interesting to include them in further work. In particular vectors in a vector-multiplet have canonical scaling dimension one, and play a central role in the construction of \cite{Samtleben:2011fj} (whereas the scalars in a linear multiplet have  scaling dimension four). In our construction we will obtain five-dimensional gauge fields, and hence non-abelian interactions, through the dimensionally reduced tensor multiplet.
 
The $(1,0)$ superalgebra admits an $SU(2)$ R-symmetry and is generated by a spinor $\epsilon_\alpha$, which is chiral and satisfies a sympletic Majorana condition
\begin{align}
\gamma_{012345}\epsilon_\alpha = \epsilon_\alpha\qquad \epsilon^\alpha = (\epsilon_\alpha)^* = i\varepsilon^{\alpha\beta}C\gamma^0\epsilon_\beta	\ .
\end{align}
Here $\alpha,\beta=1,2$ label the $SU(2)$ R-symmetry and $C$ is the unitary charge conjugation matrix $C\gamma_\mu C^{-1} = -\gamma_\mu^T$, $C^T=C$ with $\mu=0,1,2,...5$. We define $\bar\psi = i\psi^\dag\gamma^0$ for any spinor $\psi$. We have also introduced a notation where complex conjugation raises (or lowers) the R-symmetry index (as well as any flavour index). 
Thus for example $\bar\epsilon^\alpha = i
\epsilon_\alpha^\dag\gamma^0$. In this section we will restrict our attention to free abelian fields in flat six-dimensional spacetime.

 
 
A hyper-multiplet consists of two complex scalars $X^\alpha$, $\alpha = 1,2$ and an anti-chiral fermion  $\chi$, $\gamma_{012345}\chi=-\chi$.  In the free, abelian theory the six-dimensional supersymmetry transformations are
\begin{equation}
	\begin{split}
	\delta X^\alpha &= -\bar\epsilon^\alpha \chi	\\
\delta\chi &= 2\gamma^\mu \partial_\mu X^{\alpha}\epsilon_\alpha - 8X^\alpha\eta_\alpha\ .
\end{split}
\end{equation}
where $\epsilon_{\alpha}$ is a conformal Killing spinor satisfying 
\begin{align}
\partial_\mu \epsilon_\alpha = \gamma_\mu\eta_\alpha	\ .
\end{align}
Reducing a hyper-multiplet on a null direction or otherwise is relatively straight-forward, as one simply demands that the fields do not depend on the relevant null coordinate and the multiplet is otherwise unchanged. 

We can combine two hyper-multiplets $(X^\alpha{}_{\dot \beta}, \chi_{\dot \beta})$ , ${\dot\beta}=1,2$, if we simultaneously impose a reality conditions
\begin{equation}
	\begin{split}\label{20reality}
		X_\alpha{}^{\dot\beta} &= (X^\alpha{}_{\dot\beta})^* = \varepsilon_   {\alpha\gamma}\varepsilon^{{\dot\beta}{\dot \delta}} X^\gamma{}_{\dot\delta}\\
	\chi^{\dot\beta} &= (\chi_{\dot\beta})^* = \varepsilon^{{\dot\beta}{\dot\delta}}\chi_{\dot\delta}	\ .
	\end{split}
\end{equation}
This leads to the so-called scalar multiplet of \cite{Bergshoeff:1985mz}. We will describe half-hyper-multiplets at the end of this section once we introduce a non-abelian structure.

Of particular interest is the tensor-multiplet, which consists of a real scalar $\phi$, a self-dual and closed three-form $H_{\mu\nu\rho}$, and anti-chiral fermions $\lambda_\alpha$: $\gamma_{012345}\lambda_\alpha=-\lambda_\alpha$.  These are also subjected to the sympletic Majorana constraint 
\begin{equation}
\lambda^\alpha = (\lambda_\alpha)^*=i\varepsilon^{\alpha\beta}C\gamma^0\lambda_\beta\ . 	
\end{equation}
The supersymmetry transformation are 
\begin{equation}
	\begin{split}
		\delta \phi & = -\bar\epsilon^\alpha\lambda_\alpha\\
		\delta H_{\mu\nu\rho} & = 3\bar\epsilon^\alpha\gamma_{[\mu\nu}\partial_{\rho]}\lambda_\alpha - 3\bar\eta^\alpha\gamma_{\mu\nu\rho}\lambda_\alpha\\
		\delta \lambda_\alpha & = \gamma^\mu\partial_\mu \phi\epsilon_\alpha + \frac{1}{2\cdot 3!}H_{\mu\nu\lambda}\gamma^{\mu\nu\lambda}\epsilon_\alpha+4\phi\eta_\alpha\ .
	\end{split}
\end{equation}
where again $\partial_\mu\epsilon_\alpha = \gamma_\mu\eta_\alpha$. These only close on-shell and with the constraint $H=\star H$.

Difficulties arise when one wants to make an interacting theory, presumably based on allowing the fields to become non-abelian. For example there is no clear understanding of a non-abelian self-dual tensor arising as a field strength of some kind of non-abelian 2-form connection (for some recent developments see \cite{Saemann:2019leg})   

One way to proceed is to reduce the theory on a spacelike circle parametrised by $x^5\cong x^5+2\pi R_5$. In this case the hyper-multiplet is essentially unchanged.  On the other hand, for a tensor multiplet, as a consequence of self-duality, the independent components of the three-form are encoded in $F=R_5i_{5}H$. One can then identify $F=dA$, where $A$ is a five-dimensional one-form gauge field. Reduction then leads to a five-dimensional vector multiplet, which one can then make non-abelian.  The resulting theories are well-studied as one can indeed construct more-or-less traditional five-dimensional gauge theory Lagrangians. However, the problem is that such theories are power-counting non-renormalizable. As a result their relation to the original superconformal field theory is at best obscured. 

Another approach is to reduce on a null circle; $x^+\cong x^+ + 2\pi R_+$. In the simplest form this leads to the so-called DLCQ constructions. More recently, as discussed above,  a variation of this has been studied where the $x^+$ direction is only conformally compactified \cite{Lambert:2019jwi} so that the non-compact six-dimensional theory can be reconstructed \cite{Lambert:2020zdc}. In these reductions one finds the following fields from the reduction of the three-form:
\begin{align}
	\begin{split}
		H_{ij+} &\sim F_{ij}\\
		H_{ij-} & \sim  G_{ij}\ ,
	\end{split}
\end{align}
where   $i,j=1,2,3,4$ and we use  $\sim$ as the exact relation depends upon the details of the null reduction \cite{Lambert:2020scy}. Furthermore  $F\sim -\star_4 F$ and $G=\star_4 G$ where $\star_4$ is the Hodge star in $x^1,...,x^4$. One finds that 
the remaining components of $H$ can be determined using six-dimensional self-duality from $F$ and $G$. It was observed in \cite{Lambert:2018lgt,Lambert:2020scy} that non-Lorentzian five-dimensional Lagrangians can be constructed and generalized to non-abelian interacting fields by taking $F=dA$, where $A=(A_-,A_i)$ is a five-dimensional gauge field  and $G=\star_4 G$ off-shell. In the action $G$  acts as a Lagrangian multiplier that imposes the anti-self-duality of $F$ on-shell. Thus after a null reduction the tensor multiplet gives rise to the following fields:
\begin{align}
(\phi, H_{\mu\nu\lambda}, \lambda_\alpha) \xrightarrow{\text{Null Reduction }} (\phi,A_-,A_i,G^+_{ij},\lambda_\alpha)\ ,
\end{align}
where the $+$-superscript  indicates that $G^+_{ij}$ is taken to be self-dual off-shell.

Our task in the next section is to construct interacting non-abelian Lagrangian gauge theories in five-dimensions that might arise from six-dimensional $(1,0)$ theories made from hyper and tensor-multiplets. As discussed above, we expect these to have 3/4 of the supersymmetry and superconformal symmetry as well as an $SU(1,3)$ conformal symmetry. We will see that this is indeed the case.

\section{Actions}

In this section we  construct non-Lorentzian five-dimensional actions and show that they admit an $SU(1,3)$ conformal symmetry, along with $4$ supersymmetries and $8$ conformal supersymmetries.  
Our approach is to take the action in \cite{Lambert:2019jwi} where the symmetries are appropriate to six-dimensional  $(2,0)$ supersymmetry and all the fields are in the adjoint representation. We then rewrite it in terms of $(1,0)$ supermultiplets (see the Appendix for this map). This allows us to generalise the resulting action to an arbitrary number of hyper-multiplets, in arbitrary representations of the gauge group.

  To this end we consider matter content of a null reduced tensor multiplet $(\phi,A_-,A_i,G^+_{ij},\lambda_\alpha)$, which takes values in the adjoint of some gauge group $\cal G$, and $K$ hyper-multiplets $(X^\alpha{}_m,\chi_{ m} )$ $m=1,...,K$, which can take values in a vector space ${\cal V}$ that carries a  representation $\Pi({\cal G})$ of the gauge group $\cal G$. In our conventions the generators of this representation are Hermitian $T_a$, $a=1,...,{\rm dim (\cal G})$ (for notational clarity we do not put an $m$-index on ${\cal V}$ and $T_a$  but it is possible that each hyper-multiplet is in a different representation). Their complex conjugates  $(X_\alpha{}^m,\chi^{ m} )$ therefore take values in the complex conjugate representation with generators $-T^*_a$. One can also think of just one hyper-multiplet but in a reducible representation of $\cal G$. We assume that the Lie-algebra has an invariant non-degenerate inner-product $(\ \ ,\ \ )$ and each representation has an invariant non-degenerate inner-product $\langle\ \ ,\ \ \rangle$. We will assume that the latter is complex anti-linear in general, and we use a notation whereby the first entry is explicitly conjugated: {\it e.g.} on ${\mathbb C}^N$ we write the inner-product as $\langle {  w}^*,{ z}\rangle =\sum_{r}{{w}^\dag}_{r} {z}_{r}$.
  
Starting with the action constructed in   \cite{Lambert:2019jwi}  we are led to propose the following action 
 \begin{align}
\begin{split}\label{action}
S = \frac{1}{{g^2_{\text{YM}}}} \int dx^{-} d^4 x \Big \{ \frac{1}{2} \big (F_{i-}, F_{i-}  \big ) + \frac{1}{2} \big (\mathcal{F}_{ij}, G^+_{ij} \big ) -\frac{1}{2} \big (\nabla_{i} \phi,\nabla_{i} \phi \big ) -  \big \langle  \nabla_i {X}_{\alpha}{}^{m},\nabla_i X^{\alpha}{}_{m}  \rangle \\
- \frac{1}{2}  \big ( \bar{\lambda}^{\alpha}, \gamma_{+} D_{-} \lambda_{\alpha} \big ) + \frac{1}{2} \big ( \bar{\lambda}^{\alpha}, \gamma_{i} \nabla_{i} \lambda_{\alpha} \big )  - \frac{1}{2} \big \langle \bar{\chi}^{m}, \gamma_{+} D_{-} \chi_{m} \big \rangle+ \frac{1}{2} \big \langle \bar{\chi}^{m}, \gamma_{i} \nabla_{i} \chi_{m} \big \rangle\\
   -  \frac{i}{2} \big ( \bar{\lambda}^{\alpha}\gamma_{+}, [ \phi, \lambda_{\alpha}] \big )+ \frac{i}{2} \big \langle \bar{\chi}^{m} \gamma_{+} , \phi (\chi_{m}) \big \rangle 
  + i\big \langle \bar{\chi}^{m} \gamma_{+} , \lambda_{\alpha} (X^{\alpha}{}_{m}) \big \rangle- i\big \langle \bar{\lambda^\alpha} (X_{\alpha}{}^{m}),\gamma_{+}  \chi_{m}\big \rangle 
\Big \}\ .
\end{split}
\end{align}
The covariant derivative is defined on a hyper-multiplet as
\begin{equation}
D_-X^\alpha{}_m = \partial_-X^\alpha{}_m	 - iA_-(X^\alpha{}_m	)\qquad D_iX^\alpha{}_m = \partial_iX^\alpha{}_m	 - iA_i(X^\alpha{}_m	)\ ,
\end{equation}
where $A_-(X^\alpha{}_m	),A_i(X^\alpha{}_m	)$ {\it etc.}, are taken to act in the appropriate representation: {\it e.g.} $\phi(X^\alpha{}_m)=\phi^aT_a(X^\alpha{}_m)$. In addition we define 
\begin{equation}
	\begin{split}
{\cal F}_{ij} &= F_{ij} - \frac12 \Omega_{[i|k}x^k F_{j]-}\\
\nabla_i & = D_i - \frac12\Omega_{ij}x^j D_-\ .	
	\end{split}
\end{equation}
Where, for the tensor multiplet, we have 
\begin{equation}
D_-\phi = \partial_-\phi -i [A_-,\phi]\ ,\qquad 	D_i\phi = \partial_i\phi -i [A_i,\phi]\ , 
\end{equation}
and similarly for the other fields.



\subsection{Supersymmetries}

We note that the $\Omega$-deformed Minkowski space (\ref{OM}) admits a maximal number of (conformal) Killing spinors
\begin{equation}
	D_\mu \epsilon_\alpha  = \gamma_\mu\eta_\alpha  \ .
\end{equation}
However only those that are independent of  $x^+$ are expected to lead to symmetries of the five-dimensional action.
Explicitly these Killing spinors are of three types \cite{Lambert:2019jwi}:
\begin{equation}
\begin{split}
\epsilon_\alpha &= {\zeta_{-\alpha}}\qquad  \eta_\alpha=0\\
\epsilon_\alpha &= {\zeta'_{+\alpha}}+\frac12 x^i \Omega_{ij}\gamma_{j}\gamma_-{\zeta'_{+\alpha}}	\qquad \eta_\alpha = \frac{1}{16} \Omega_{ij}\gamma_{ij}\gamma_-{\zeta'_{+\alpha}}\\
\epsilon_\alpha & = -\frac12 x^i  \gamma_{i}\gamma_+{\zeta''_{-\alpha}}
 -\frac14 \Omega_{ik} \gamma_{kj} x^ix^j{\zeta''_{-\alpha}}+x^-{\zeta''_{-\alpha}}\qquad \eta_\alpha =-\frac12 \gamma_+{\zeta''_{-\alpha}}-\frac{1}{16} \Omega_{ij}\gamma_{ij}x^k\gamma_k\zeta''_{-\alpha}\ ,
  \end{split}
\end{equation}
 where $\zeta_{-\alpha}, \zeta'_{+\alpha},\zeta''_{-\alpha}$ are constant spinors. Here the $\pm$ subscript refers to their  $\gamma_{-+}$ chirality. Thus each one has four real independent components.

The action (\ref{action}) is indeed supersymmetric with respect to these spinors. In particular the transformations of the hyper-multiplets are 
\begin{align}
\begin{split}  
\delta X^\alpha{}_{m}&=  -\bar\epsilon^{\alpha}\chi_m\\  
\delta \chi_{m} &= -2\gamma_+ D_{-} X^\alpha{}_{m}  \epsilon_{\alpha} + 2\nabla_i X^{\alpha}{}_m \gamma_i \epsilon_{\alpha} + 2i  \phi(X^{\alpha}{}_m)\gamma_{+} \epsilon_{\alpha}-8X^\alpha{}_m\eta_{\alpha} \ ,
\end{split}
\end{align}
and those of the tensor multiplet are
\begin{align}
\begin{split}
    &\delta \phi = - \bar{\epsilon}^{\, \alpha}\lambda_{\alpha} \\
    &\delta A_{-} =  - \bar{\epsilon}^{\, \alpha} \gamma_{+-}\lambda_{\alpha} \\
    &\delta A_{i} = -\bar{\epsilon}^{\, \alpha} \big( \gamma_{+i}+\frac12 \Omega_{ij}x^j\gamma_{+-} \big) \lambda_{\alpha} \\
    &\delta G^+_{ij} = - \frac{1}{2}  \bar{\epsilon}^{\, \alpha} \gamma_{+} \gamma_{-} \gamma_{ij} D_{-} \lambda_{\alpha}+\frac{1}{2}  \bar{\epsilon}^{\, \alpha} \gamma_{k} \gamma_{ij} \gamma_{-} \nabla_{k} \lambda_{\alpha}  + \frac{i}{2}\bar{\epsilon}^{\, \alpha} \gamma_{+}\gamma_{-} \gamma_{ij} [\phi, \lambda_{\alpha}]\\
    &\qquad \quad -\frac{1}{2}\bar{\epsilon}^{\, \alpha} \gamma_{+} \gamma_{-} \gamma_{ij}   [[X_{\alpha}{}^{m}, \chi_{m}]] -\frac{1}{2} [[ \bar \chi^{m},X^{\alpha}{}_{m}]] \gamma_{-} \gamma_{+} \gamma_{ij}  \epsilon_\alpha -3\bar\eta^\alpha\gamma_-\gamma_{ij}\lambda_\alpha\\ 
   &\delta \lambda_{\alpha} = - F_{-i}\gamma_{+-}\gamma_{i}\epsilon_{\alpha} - \frac{1}{4}\mathcal{F}_{ij} \gamma_-\gamma_{ij} \epsilon_{\alpha} -\frac{1}{4}G_{ij}\gamma_+\gamma_{ij} \epsilon_{\alpha} -   \gamma_+D_{-} \phi \epsilon_{\alpha} + \gamma_{i}\nabla_{i}\phi \epsilon_{\alpha} \\
    &\qquad \quad +[[ X_{\alpha}{}^m,X^{\beta}{}_m]]\gamma_{+}\epsilon_{\beta} 
    -\varepsilon_{\alpha\gamma}\varepsilon^{\beta\delta}[[ X_{\delta}{}^m,X^{\gamma}{}_m]] \gamma_{+}\epsilon_{\beta}   
    +4\phi\eta_\alpha\ .
\end{split}
\end{align} 
Here we have introduced a map: 
 \begin{equation}
 	[[\ ;\ ]]:{\cal V}^*\times {\cal V} \to \rm Lie( \mathcal{G})\ ,
 \end{equation} 
 which is defined to satisfy, for any $X,Y\in {\cal V}$ and $\phi\in {\rm Lie}(\cal G)$,
 \begin{equation}\label{compatible}
([[X^*,Y]],\phi) = i\langle X^*,\phi(Y) \rangle 	\ .
 \end{equation}
In terms of components this tells us that
\begin{align}
[[X^*;Y]]^a
= i\langle X^* ,T_b(Y)\rangle \kappa^{ab}  
\ ,
\end{align} 
where $\kappa^{ab}$ is the inverse inner-product on ${\rm Lie}(\cal G)$ evaluated on the adjoint basis: $\kappa_{ab} = (T^{\text{adj}}_a,T^{\text{adj}}_b)$.
With our conventions 
\begin{align}
\left([[X^*;Y]]^a\right)^* = - [[Y^*;X]]	^a\ .
\end{align} 
The normalisation is chosen so that, in the special case that the fields are real and take values in the adjoint representation, $[[X,Y]]=-i[X,Y]$ and in this case the condition (\ref{compatible}) just asserts the invariance of the inner-product under the adjoint action.

\subsection{Conformal Symmetries}
 
Six-dimensional Minkowski space admits a family of 28 linearly independent conformal Killing vectors that obey  
\begin{align}
	D_{\mu}k_{\nu} + D_{\nu}k_{\mu} = \omega g_{\mu \nu}\ . 
\end{align}
A six-dimensional conformal field theory inherits an $SO(2,6)$ spacetime symmetry group. The same is true for the conformally equivalent spacetime (\ref{OM}). However, as was shown in \cite{Lambert:2019fne}, demanding that the fields are independent of $x^+$  reduces the $SO(2,6)$ conformal group to $SU(1,3)$. In terms of an $AdS_7$ dual description the $SO(2,6)$ symmetry group  is broken to $SU(1,3)$ by a timelike reduction \cite{Pope:1999xg}. For the presently considered spacetime, one finds the general form for such a $k^{\mu}$ and $\omega$ can be divided into seven classes: 
\begin{align}
 &{\rm type}\ I \ \ \ \  \left(b,0,0,0,0,0\right)\nonumber\\
&{\rm type}\ II \ \ \ \left(0,c,0,0,0,0\right)\nonumber\\
&{\rm type}\ III\ \   \left(0,\frac12 c_i\Omega_{ij}x^j,c_i\right)\nonumber\\
&{\rm type}\ IV \ \ \   \left(0,0,M_{ij}x^j\right)\\
&{\rm type}\ V \ \ \ \   \left(0,\omega_1 x^-,\frac12 \omega_1x_i\right)\nonumber\\
&{\rm type}\ VI \  \left(v_ix^i,\frac12 x^-v_i\Omega_{ij}x^j - \frac18 R^{-2} |x|^2x^kv_k,x^-v_i - \frac12 x^kv_k\Omega_{li}x^l +\frac12 v^kx^l\Omega_{kl}x^i-\frac14|x|^2v^k\Omega_{ki}\right)\nonumber\\
&{\rm type}\ 
VII \    \left(\frac14 \omega_2|x|^2,\frac12\omega_2(x^-)^2- \frac{\omega_2}{32} R^{-2} |x|^4,-\frac{1}{8}\omega_2\Omega_{ki}|x|^2x_k+ \frac12\omega_2x_ix^-\right)\nonumber\ .
\end{align}
where the conformal factor is given by
\begin{equation}
	\omega = \omega_1+v_i\Omega_{ij}x^j + \omega_2 x^-\ .
\end{equation}
Here $b,c,c_i,v_i,\omega_1,\omega_2$ are constants and $M_{ij}$ is an anti-symmetric matrix that commutes with $\Omega_{ij}$. In \cite{Lambert:2019fne} it was shown types II - VII form a representation of $SU(1,3)$ that the $(2,0)$ version of the action (\ref{action}) admits these as spacetime symmetries. We now want to extend this result to the more general $(1,0)$ theories. 

Cases I -  IV are rather obvious spacetime symmetries
\begin{enumerate}[I]
\item - Translations in $x^{+}$, which are trivial in our five-dimensional action.
\item - Translations in $x^{-}$. 
\item - Translations in $x^{i}$, with  a compensating shift in $x^{-}$.
\item - Spatial rotations that preserve $\Omega$.
\end{enumerate}

Furthermore it is not hard to identify case V as the  Liftshitz rescaling 
\begin{align}
x^{-} \longrightarrow \zeta x^{-}, \quad x^{i} \longrightarrow \zeta^{1/2} x^i,
\end{align}
with the fields in the tensor and hyper multiplets transforming as
\begin{align}
\begin{split}
\phi &\longrightarrow \zeta^{-1} \phi  \\
{\lambda_{+}}_{\alpha} &\longrightarrow \zeta^{-3/2} {\lambda_{+}}_{\alpha}  \\ 
{\lambda_{-}}_{\alpha} &\longrightarrow \zeta^{-1} {\lambda_{-}}_{\alpha} \\
A_{-} &\longrightarrow \zeta^{-1} A_{-}  \\ 
A_{i} &\longrightarrow \zeta^{1/2} A_{i}   \\ 
G_{ij} &\longrightarrow \zeta^{-2} G_{ij}
\end{split}
\begin{split}
&X^{\alpha}_{m} \longrightarrow \zeta^{-1} X^{\alpha}_{m} \\
&{\chi_{+}}_{m} \longrightarrow \zeta^{-3/2} {\chi_{+}}_{m} \\
&{\chi_{-}}_{m} \longrightarrow \zeta^{-1} {\chi_{-}}_{m} \\
& \hfill \\
& \hfill \\
& \hfill 
\end{split}
\end{align}
Here the subscript $\pm$ on spinors indicates eigenvalue under $\gamma_{-+}$. 
 
 The remaining transformations are not so obviously symmetries, they correspond to special conformal transformations.
 For type VI the infinitesimal coordinate transformations are
\begin{align}
\delta_{} x^-&=\frac{1}{2} \Omega_{ij} v_i x^j x^- - \frac{1}{8} R^{-2} |x|^2 v_i x^i \nonumber\\
 \delta_{} x^i  &= \frac{1}{2}\Omega_{jk}v_j x^k x^i + v_i x^- + \frac{1}{2}\Omega_{ij}v_k x^j x^k + \frac{1}{4} |x|^2 \Omega_{ij} v_j\ ,
\end{align}
which is found to be a symmetry of the action if the tensor multiplet transforms as
\begin{align}
\begin{split}
\delta \phi &= - \Omega_{ij} v_{i} x^{j} \phi \\
\delta {\lambda_{+}}_{\alpha} &= - \frac{3}{2} \Omega_{ij} v_i x^j {\lambda_{+}}_{\alpha} + \frac{1}{2} v_i \Gamma_{+} \Gamma_{i} {\lambda_{-}}_{\alpha} + \frac{1}{4} \Lambda_{ij} \Gamma_{ij} {\lambda_{+}}_{\alpha} \\
\delta {\lambda_{-}}_{\alpha} &= - \Omega_{ij} v_i x^j {\lambda_{-}}_{\alpha} + \frac{1}{4} \Lambda_{ij} \Gamma_{ij} {\lambda_{-}}_{\alpha} \\
\delta A_{-} &= -\frac{1}{2} \Omega_{ij} v_i x^j A_-  - v_i A_i \\
\delta A_i &= - \frac{1}{2} \Omega_{jk} v_j x^k A_i + \frac{1}{2} ( \Omega_{ij} v_k x^k + \Omega_{ik} v_k x^j-\Omega_{jk}( v_i x^k + v_k x^i ) )A_j \\ & \quad + \frac{1}{8} ( R^{-2} |x|^2 v_i + 2R^{-2} v_j x^j x^i + 4 \Omega_{ij} v_j x^- ) A_- \\
\delta G_{ij} &=  -2\Omega_{kl} v_k x^l G_{ij}  - \frac{1}{2}( \Lambda^{ki} G_{kj} - \Lambda^{kj} G_{ki} +\varepsilon_{ijkl} \Lambda^{mk} G_{ml}) \\ & \quad + v_i F_{-j} - v_j F_{-i} + \varepsilon_{ijkl} v_k F_{-l} \ ,
\end{split}
\end{align}
and the hyper multiplets as
\begin{align}
\begin{split}
\delta X^{\alpha}_{m} &= - \Omega_{ij} v_{i} x^{j} X^{\alpha}_{m} \\
\delta {\chi_{+}}_{m} &=  - \frac{3}{2} \Omega_{ij} v_i x^j {\chi_{+}}_{m} + \frac{1}{2} v_i \Gamma_{+} \Gamma_{i} {\chi_{-}}_{m} + \frac{1}{4} \Lambda_{ij} \Gamma_{ij} {\chi_{+}}_{m} \\
\delta {\chi_{-}}_{m} &= - \Omega_{ij} v_i x^j {\chi_{-}}_{m} + \frac{1}{4} \Lambda_{ij} \Gamma_{ij} {\chi_{-}}_{m} \ .          
\end{split}
\end{align}
We have defined 
\begin{align}
\Lambda_{ij} = \frac{1}{2} ( \Omega_{ij} v_k x^k + \Omega_{ik} v_k x^j - \Omega_{jk} v_k x^i +\Omega_{ik} v_j x^k - \Omega_{jk} v_i x^k )\ .	
\end{align}
Note that in the limit that $R\to\infty$ we have $ \Omega_{ij}\to 0$, and these transformations become Galilean boosts: $\delta x^-=0,\delta x^i=v^ix^-$.

While for type VII, the infinitesimal coordinate transformations are
\begin{align}
  \delta_{}  x^- &= \frac{1}{2} \omega_2 ( x^-)^2 - \frac{1}{32}\omega_2 R^{-2} |x|^4\nonumber\\
  \delta_{}  x^i &= \frac{1}{8} \omega_2 \Omega_{ij} |x|^2 x^j + \frac{1}{2} \omega_2 x^i x^-\ .
\end{align}
which we find to be a symmetry of the action if the corresponding transformations  are
\begin{align}
\begin{split}
\delta \phi &= - \omega_{2} x^{-} \phi \\
\delta {\lambda_{+}}_{\alpha} &= - \frac{3}{2} \omega_{2} x^{-} {\lambda_{+}}_{\alpha} + \frac{1}{4} \omega_{2} x^{i} \Gamma_{+} \Gamma_{i} {\lambda_{-}}_{\alpha} + \frac{1}{4} \Lambda_{ij} \Gamma_{ij} {\lambda_{+}}_{\alpha} \\
\delta {\lambda_{-}}_{\alpha} &= - \omega_{2} x^{-} {\lambda_{-}}_{\alpha} + \frac{1}{4} \Lambda_{ij} \Gamma_{ij} {\lambda_{-}}_{\alpha} \\
\delta A_{-} &= -\omega_{2} x^{-} A_-  - v_i A_i \\
\delta A_i &= - \frac{1}{2} \omega_{2} x^{-} A_i + \frac{1}{8} \omega_2 R^{-2} |x|^2 x^i A_- + \frac{1}{8} \omega_2 ( \Omega_{ij} |x|^2 - 2\Omega_{jk} x^i x^k ) A_j \\
\delta G_{ij} &=  -2\omega_{2} x^{-} G_{ij}  - \frac{1}{2}( \Lambda^{ki} G_{kj} - \Lambda^{kj} G_{ki} +\varepsilon_{ijkl} \Lambda^{mk} G_{ml}) \\ & \quad + \frac{1}{2} \omega_{2} ( x^i F_{-j} - x^j F_{-i} + \varepsilon_{ijkl} x^j F_{-l}) \ ,
\end{split}
\end{align}
and
\begin{align}
\begin{split}
\delta X^{\alpha}_{m} &= - \omega_{2} x^{-} X^{\alpha}_{m} \\
\delta {\chi_{+}}_{m} &=  - \frac{3}{2} \omega_{2} x^{-} {\chi_{+}}_{m} + \frac{1}{4} \omega_{2} x^{i} \Gamma_{+} \Gamma_{i} {\chi_{-}}_{m} + \frac{1}{4} \Lambda_{ij} \Gamma_{ij} {\chi_{+}}_{m} \\
\delta {\chi_{-}}_{m} &= - \omega_{2} x^{-} {\chi_{-}}_{m} + \frac{1}{4} \Lambda_{ij} \Gamma_{ij} {\chi_{-}}_{m} \ .          
\end{split}
\end{align}
Where now $\Lambda$ is given by 
\begin{align}
\Lambda^{ij}=\frac{1}{4} \omega_2 \left( \Omega_{ik}x^k x^j - \Omega_{jk}x^k x^i \right) +\frac{1}{8}\omega_2\Omega_{ij}|x|^2 	\ .
\end{align}

\subsection{Enhancement to $(2,0)$ and Other Variations}

Let us now outline how a $(2,0)$ tensor multiplet, and corresponding action, can be obtained (see the Appendix for additional details). Since we need a total of five scalars one might think that it is sufficient to consider one tensor and one hyper-multiplet. However this will not work as the R-symmetry remains only $SU(2)$, and not $SO(5)$, as is required. Instead one must take a $(1,0)$ tensor multiplet $(\phi,H_{\mu\nu\lambda},\lambda_\alpha)$ along with two copies of the $(1,0)$ hyper-multiplet $(X^\alpha{}_m,\chi_m)$,  where $m=1,2$ which we combine into the scalar multiplet using the reality condition (\ref{20reality}). Crucially the representation of all multiplets must be the adjoint.  The various reality conditions reduce the total number of scalars to five, $(\phi, X^{\alpha}{}_{m})$ and the total number of on-shell fermionic degrees of freedom to eight.  

In addition to the original $SU(2)$ R-symmetry for the scalars, we find an $Sp(1)\cong SU(2)$ flavour symmetry. The resulting $SU(2)\times SU(2)\cong SO(4)$ then enhances to an $SO(5)\cong Usp(4)$ R-symmetry with the addition of $\phi$ from the $(1,0)$ tensor multiplet, this is the appropriate R-symmetry for $(2,0)$ supersymmetry. The fermions also combine to give $\psi_{r}$ where r covers the range of $m, \alpha$ which is $1,\dots,4$. In this way we can consider $\lambda_{\alpha}, \chi_{m}$ as the Weyl spinors of the Dirac spinor $\psi_{r}$
\begin{align}
\begin{rcases}
\text{$(1,0)$ tensor}  & (\phi,H,\lambda_\alpha) \\
\text{$(1,0)$ scalar}  & (X^{\alpha}{}_{m}, \chi_{m}) 
\end{rcases} \longrightarrow \ \text{$(2,0)$ tensor} \ \ (X^{r s}, \psi_r, H)\ ,
\end{align} 
with the scalars now written as transforming in the $\mathbf{5}$ of $Usp(4) \cong SO(5)$. In this way the $(2,0)$ model obtained in \cite{Lambert:2019jwi} arises from (\ref{action}) by taking two hypermultiplets in the adjoint representation and imposing the additional reality condition (\ref{20reality}). This relation is explored in more detail in the appendix, where it is performed explicitly, relating this theory to the natural M5 picture in M-theory.

More generally, if we consider a  $2L$-dimensional subset of  hyper-multiplets in the same representation $\Pi({\cal G})$ then we can impose a similar reality constraint. For this to make sense, we require that the representation is real in the sense that there exists a unitary matrix $U$ such that
 \begin{align}\label{U}
 -T_a^* = UT_aU^{-1}\ .	
 \end{align} 
This is trivially true for the adjoint representation, $U = I$. In the more general case we impose 
 \begin{equation}
 	X_\alpha{}^m = \varepsilon_{\alpha\beta}\omega^{mn}U(X^\beta{}_n)\qquad \chi^m=\omega^{mn}U(\chi_n)\ ,
 \end{equation}
 where $\omega^{mn}$ is an anti-symmetric matrix which squares to minus the identity. This reduces the flavour symmetry from $U(2L)$ to $Sp(L)$. However unlike the $(2,0)$ case discussed above there will not be, in general, an enhancement of supersymmetry.

Lastly we mention half-hyper-multiplets. These can be defined whenever a  hyper-multiplet take values in a pseudo real-representation of the gauge group. In this case the unitary matrix $U$ in (\ref{U}) satisfies 
\begin{align}
U^{-1}=-U^*\ ,
\end{align}
or equivalently $U^T=-U$. This implies that the dimension of the representation must be even.
This enables us to impose the  conditions
\begin{align}
	X_\alpha = \varepsilon_{\alpha\beta}U(X^\beta)\qquad \chi^* =Ci\gamma^0U(\chi)\ .
\end{align}

 \section{Discussion and Conclusion}
 
 The main result of this paper was the construction of five-dimensional non-abelian gauge theories, given in (\ref{action}),  without Lorentz invariance but with an $SU(1,3)$ spacetime conformal symmetry.  In addition these actions admit 4 supersymmetries and 8 superconformal symmetries.  These theories also have a topological $U(1)$ current
 \begin{align}
 J_- = 	 \frac1{32\pi^2}\varepsilon_{ijkl} {\rm tr} (F_{ij} F_{kl})\qquad J_i  =\frac1{16\pi^2}\varepsilon_{ijkl}  {\rm tr} (F_{-j}F_{kl})\ .		
 \end{align}
 As is the case for the $(2,0)$ theory \cite{Lambert:2019jwi} we expect that one can construct gravitational duals  in the form of a $U(1)$ fibre reduction of $AdS_7$ \cite{Pope:1999xg} but now starting with the solutions obtained in \cite{Apruzzi:2013yva,Apruzzi:2015wna}.
 
 We hope these theories are of interest but our primary motivation is that they can be identified with Lorentz invariant six-dimensional $(1,0)$ conformal field theories  on conformally compactified Minkowski space (\ref{OM}). The Kaluza-Klein mode is then identified with the charge arising from the topological $U(1)$ current. Using the techniques developed in \cite{Lambert:2020zdc} we hope that these Lagrangian field theories can be used to describe six-dimensional Lorentzian conformal field theories.  In contrast to other studies, our actions are based on a null reduction of six-dimensional hyper and tensor multiplets with no vector multiplets. It could be of interest to find more general theories by including other multiplets.

 We can also take $R\to\infty$ which sets $\Omega_{ij}=0$. In this case there is no conformal compactification and we are simply reducing over $x^+$ which we impose by hand to be compact with period $2\pi R_{+}$. This is the DLCQ description.  Furthermore the type VI symmetry above becomes a boost.  In our Lagrangians setting $\Omega_{ij}=0$ means that ${\cal F}_{ij}=F_{ij}$ and the constraint imposed by $G^+_{ij}$ is simply that $F_{ij}$ is anti-self-dual. Thus the dynamics is reduced to motion on instanton moduli space. This in line with a well-known prescription \cite{Aharony:1997pm,Aharony:1997an}.  However our models allow one to understand in a controlled way how various aspects of the six-dimensional theory are encoded in the resulting quantum mechanics in terms of additional parameters and fields, for example see \cite{Mouland:2019zjr}. 
 
 Finally, let us recall that in the case of the $(2,0)$ theory the   action can be obtained by a non-Lorentzian rescaling  of five-dimensional  maximally supersymmetric   Yang-Mills theory \cite{Lambert:2019nti,Lambert:2019jwi}. The Liftshitz scaling then arises as a symmetry of the fixed point.  For $\Omega_{ij}\ne 0 $ there is a dual AdS description where the Yang-Mills theory arises from  M5-branes reduced on a timelike $S^1$ fibration of $AdS_7$. The $SU(1,3)$ conformal symmetry, including a  Liftshitz scaling,  emerges in the limit where the M5-brane is taken to the boundary of $AdS_7$ \cite{Lambert:2019jwi}. We expect analogous constructions to hold for the case of the $(1,0)$ theories we have constructed here.

\section*{Acknowledgements}

We would like to thank R. Mouland and M. Trepanier for helpful discussions. N.L. was supported in part by STFC grant ST/L000326/1 and T.O. by the STFC studentship ST/S505468/1.

\section*{Appendix A:  Relation to $Spin(1,10)$}
 
 Let us start with the $32\times 32$ $\Gamma$-matrices of $Spin(1,10)$. In this Appendix we use a hat to denote elements associated to $Spin(1,10)$ and $M=0,1,2,...,10$. These have a charge conjugation matrix $\hat C$:
 \begin{align}
\label{eq:11dCliff}
\hat{C}^T = - \hat{C}, \qquad (\hat{C} \hat{\Gamma}^{M})^T = \hat{C}\hat{\Gamma}^{M} \ .
\end{align}
The spinors in eleven-dimensions  satisfy a Majorana condition
\begin{align}
\label{eq:11dMaj}
\bar{\hat{\psi}} := \hat{\psi}^T \hat{C} = \hat{\psi}^{\dagger} i\hat{\Gamma}^0 \ ,
\end{align}
and similarly for the supersymmetry generator $\hat\epsilon$. We  decompose the $\Gamma$ and charge conjugation matrices as ($\mu=0,1,2,...,5$, $A=6,7,8,9$)
\begin{align}
\begin{split}
\hat{\Gamma}^{\mu} &= \gamma^{\mu} \otimes \mathbb{I}_4 \\
\hat{\Gamma}^{A} &= \gamma_{*} \otimes \rho^{A}  \\
\hat{\Gamma}^{10} &= \gamma_{*} \otimes \rho_{*} \ .
\end{split}
\end{align}
Here $\gamma_* = \gamma_0\gamma_1...\gamma_5$, $\rho_* = \rho_6\rho_7\rho_8\rho_9$ and 
\begin{align}
\hat{C} = C \otimes C_4 \ .
\end{align}
Applying this decomposition to eq (\ref{eq:11dCliff}), we see the inherited conditions 
\begin{align}
\begin{split}
\label{eq:6,4dCliff}
C^{T} = C, \quad (C \gamma^{\mu})^T = - C \gamma^{\mu} \\
C^{T}_{4} = -C_{4}, \quad (C_4 \rho^A)^T = - C_4 \rho^A  \ .
\end{split}
\end{align}
We can combine the first of each of these with the general unitarity of charge conjugation matrices to obtain
\begin{align}
\label{eq:inverses}
\begin{split}
&C^{*} = C^{-1}, \qquad  \qquad \ \; C^{*}_4 = - C^{-1}_4, \\
&{\gamma^{\mu}}^T = - C\gamma^{\mu} C^{-1}, \quad {\rho^A}^T = C_4 \rho^A C^{-1}_4 \ .
\end{split}
\end{align}

Next we must expand the eleven-dimensional spinors in a basis $(\xi^\alpha_+, \xi^{\dot \alpha}_-)$
 of the internal spinor space $Spin(4)$:
\begin{equation}
\begin{split}
	\hat\psi &=\sum_{\alpha}\psi_{\alpha}\otimes \xi^{\alpha}_+	+ \sum_{\dot\alpha}\psi_{\dot\alpha}\otimes \xi^{\dot\alpha}	_- \\
	\hat\epsilon & =  \sum_{\alpha} \epsilon_{\alpha}\otimes \xi^{\alpha}_++\sum_{\dot \alpha} \epsilon_{\dot\alpha}\otimes \xi^{\dot\alpha}_-\ .
\end{split}
\end{equation}
Here we have split the basis according to their internal chirality $\rho_*\xi^\alpha_+ = \xi^\alpha_+$, $\rho_*\xi^{\dot \alpha}_- = -\xi^{\dot \alpha}_-$ with $\alpha,\dot\alpha=1,2$.\footnote{In the main body of this paper a $\pm$ subscript on a six-dimensional spinor refers to its chirality under $\gamma_{05}$.} We will also assume that these have been normalised to 
\begin{equation}
	\begin{split}
(\xi_+^\beta)^\dag \xi^\alpha_+ &=\delta^\alpha_\beta\\	
(\xi_-^{\dot\beta})^\dag \xi^{\dot \alpha}_- &=\delta^{\dot \alpha}_{\dot\beta}\\	
(\xi_-^{\dot \beta})^\dag \xi^\alpha_+ &=0\ .
	\end{split}
\end{equation}

The condition of $(2,0)$ supersymmetry imposes the constraints $\hat\Gamma_{012345}\hat\psi = -\hat\psi$ and $\hat\Gamma_{012345}\hat\epsilon= \hat\epsilon$ which in turn imply
\begin{equation}
	\begin{split}
	\gamma_*\psi_\alpha &= - 	\psi_\alpha\qquad \gamma_*\psi_{\dot \alpha} = - 	\psi_{\dot \alpha} \\
	 \gamma_*\epsilon_\alpha &=  	\epsilon_\alpha\qquad \gamma_*\epsilon_{\dot \alpha} =  	\epsilon_{\dot \alpha}\ .
	\end{split}
\end{equation}
In six-dimensions we do not have Majorana spinors, but symplectic Majorana-Weyl ones. Thus the decomposition of the eleven-dimensional Majorana condition (\ref{eq:11dMaj}) is solved by the conditions\begin{align}
\begin{split}
(\psi_{\alpha})^* &= \varepsilon^{\alpha \beta} C i\gamma^{0} \psi_{\beta} \\
(\epsilon_{\alpha})^* &= \varepsilon^{\alpha \beta} Ci\gamma^{0} \epsilon_{\beta} \\
(\xi^{\alpha}_+)^* &= \varepsilon_{\alpha \beta} C_4 \xi^{\beta} _+\ ,
\end{split}
\end{align}
and similarly for $\psi_{\dot\alpha}$, $\epsilon_{\dot\alpha}$ and $\xi^{\dot\alpha}_-$.\footnote{In the main body of this paper we use a notation where complex conjugation raises/lowers $\alpha$ indices.
}
 Here $\varepsilon_{\alpha\beta}=\varepsilon^{\alpha\beta}$ is the Levi-Civita symbol.

Next we need to impose a further constraint to reduce to $(1,0)$ supersymmetry. For this we project to
\begin{equation}
\hat\Gamma^{10}\hat\epsilon = \hat \epsilon\ .	
\end{equation} 
 We further identify $\hat\lambda = \frac12 (1-\hat\Gamma^{10})\hat \psi$ and  $\hat\chi = \frac12 (1+\hat\Gamma^{10})\hat \chi$. Thus we must expand
\begin{equation}
\begin{split}
		\hat\epsilon & =  \sum_\alpha \epsilon_\alpha\otimes \xi^\alpha_+\\
		\hat\lambda &= \sum_\alpha \lambda_\alpha\otimes \xi^\alpha_+	\\
\hat\chi &= \sum_{\dot \alpha} \chi_{\dot \alpha}\otimes \xi^{\dot \alpha}_-	\ .
\end{split}
\end{equation}
As before we still have $\gamma_{012345}\lambda_\alpha =- \lambda_\alpha$, $\gamma_{012345}\chi_{\dot \alpha} =- \chi_{\dot \alpha}$ and $\gamma_{012345}\epsilon_\alpha = \epsilon_\alpha$.

In addition to the fermions $\hat\psi$ the $(2,0)$ representation has five real scalars $X^A,X^{10}$ and a self-dual tensor $H_{\mu\nu\lambda}$. The transformation rules are
\begin{equation}
	\begin{split}
	\delta X^A & = \bar{\hat\epsilon}\hat\Gamma^A\hat\psi\\
	\delta X^{10} & = 	\bar{\hat\epsilon}\hat\Gamma^{10}\hat\psi\\
	\delta H_{\mu\nu\rho} & = 3\bar{\hat\epsilon} \hat \Gamma_{[\mu\nu}\partial_{\rho]}\hat\psi\\
	\delta \hat \psi & = \hat\Gamma^\mu\hat\Gamma^A\partial_\mu X^A\hat\epsilon + \hat\Gamma^\mu\hat\Gamma^{10}\partial_\mu X^{10}\hat\epsilon + \frac{1}{2\cdot 3!}\hat\Gamma^{\mu\nu\rho}H_{\mu\nu\rho}\hat\epsilon\ .
	\end{split}
\end{equation}
As can be seen, selecting a special $\gamma$ matrix, $\gamma_{10}$ splits the scalar fields $I \longrightarrow (A, 10), \ A \in \{6,\dots,9\}$.  Substituting in the above decomposition one readily sees that $(X^{10},H_{\mu\nu\lambda},\lambda_\alpha)$ form a $(1,0)$ tensor multiplet, we then label $X^{10} = \phi$

To untangle the remaining fields $(X^A,\chi_{\dot\alpha})$ we note that we can define
\begin{equation}
\begin{split}
T^{A\beta}{}_{\dot\alpha} = (\xi^{\dot\alpha}_-)^\dag\rho^A\xi^{\beta}_+	\ ,
\end{split}	
\end{equation}
which satisfies
\begin{equation}
\begin{split}
T^{A\beta}{}_{\dot\alpha} (T^{A\delta}{}_{\dot\gamma})^*=  2\delta^{\dot\gamma}_{\dot\alpha}\delta^\beta_{\delta}\ .  
\end{split}	
\end{equation}
This allows us to define
\begin{equation}
\begin{split}
X^\beta{}_{\dot \alpha} &= \frac12 X^AT^{A\beta}{}_{\dot\alpha}\\
X^A &= (T^{A\alpha}{}_{\dot\beta})^*X^\alpha{}_{\dot\beta}
\  ,
\end{split}	
\end{equation}
and the reality of $X^A$ implies that
\begin{equation}
\begin{split}\label{Xreal}
(X^\beta{}_{\dot \alpha})^* = \varepsilon_{\beta\alpha}\varepsilon^{\dot\alpha\dot\beta} X^\alpha{}_{\dot\beta} \  .
\end{split}	
\end{equation}
We then recognise $(X^\beta{}_{\dot \alpha},\chi_{\dot\alpha})$ as a doublet of hyper-multiplets $\dot\alpha=1,2$ subjected to the additional reality constraint (\ref{Xreal}).

\bibliographystyle{JHEP}
\bibliography{Null10}

\end{document}